\documentclass[12pt]{article}

\usepackage{cite}
\usepackage{dsfont}
\usepackage{amsfonts}
\usepackage{amssymb,latexsym}
\usepackage{amsmath,amsbsy}
\usepackage{epsfig,bm}
\usepackage{graphicx,comment}
\usepackage[active]{srcltx}

\newcommand{\be}{\begin{equation}}
\newcommand{\ee}{\end{equation}}
\newcommand{\bea}{\begin{eqnarray}}
\newcommand{\eea}{\end{eqnarray}}
\newcommand{\bem}{\begin{multline}}
\newcommand{\eem}{\end{multline}}
\newcommand{\beg}{\begin{gather}}
\newcommand{\eeg}{\end{gather}}

\newcommand{\as}{\alpha_s}

\def\eq#1{{Eq.~(\ref{#1})}}
\def\fig#1{{Fig.~\ref{#1}}}
\newcommand{\ben}{\begin{eqnarray*}}
\newcommand{\een}{\end{eqnarray*}}

\newcommand{\amu}{\alpha_\mu}

\newcommand{\pd}{\partial}

\def\peq#1{{(\ref{#1})}}

\graphicspath{{Figs/}}

\begin{document}

\begin{center}
  {\bf \Large Running Coupling Corrections  \\[5mm] to Nonlinear Evolution for Diffractive Dissociation} \\[1cm]
  Yuri V.\ Kovchegov\footnote{kovchegov.1@asc.ohio-state.edu} \\[3mm]
  {\it\small Department of Physics, The Ohio State University, 191
    West Woodruff Avenue, \\ Columbus, OH 43210, USA}
\end{center}

\begin{abstract}
  We determine running coupling corrections to the kernel of the
  non-linear evolution equation for the cross section of single
  diffractive dissociation in high energy DIS. The running coupling
  kernel for diffractive evolution is found to be exactly the same as
  the kernel of the rcBK evolution equation.
  ~
\end{abstract}


\section{Introduction} 

In recent years running coupling corrections have been calculated for
a variety of small-$x$ observables. The running coupling corrections
for the linear Balitsky-Fadin-Kuraev-Lipatov (BFKL) evolution equation
\cite{Bal-Lip,Kuraev:1977fs} were found in \cite{Kovchegov:2006wf}
(see also \cite{Braun:1994mw,Levin:1994di}), while the corrections for
the non-linear Balitsky--Kovchegov (BK)
\cite{Balitsky:1996ub,Balitsky:1997mk,Balitsky:1998ya,Kovchegov:1999yj,
  Kovchegov:1999ua} and
Jalilian-Marian--Iancu--McLerran--Weigert--Leonidov--Kovner (JIMWLK)
\cite{Jalilian-Marian:1997jx, Jalilian-Marian:1997gr,
  Jalilian-Marian:1997dw, Jalilian-Marian:1998cb, Kovner:2000pt,
  Weigert:2000gi, Iancu:2000hn, Ferreiro:2001qy} evolution equations
were found in
\cite{Kovchegov:2006vj,Balitsky:2006wa,Gardi:2006rp,Albacete:2007yr}.
The question of the running coupling scale in gluon production was
first addressed in \cite{Kovchegov:2007vf}, with the corrections to
the lowest-order gluon production cross section calculated in
\cite{Horowitz:2010yg}.

Calculations of running coupling corrections for these small-$x$
observables serve two related purposes. One is to answer the
theoretical question about how running coupling effects enter
small-$x$ evolution and particle production. The other reason for
calculating running coupling corrections is phenomenological: it is
known that running coupling effects reduce the effective value of the
BFKL pomeron intercept
\cite{Levin:1994di,Albacete:2007yr,Albacete:2004gw} as compared to the
fixed-coupling results, leading to a slower growth with energy of
cross sections and structure functions, which is more in line with the
experimental data. This feature resulted in a very successful
phenomenology developed over the past few years for the deep inelastic
scattering (DIS) \cite{Albacete:2009fh,Albacete:2010sy} and for heavy
ion collisions \cite{ALbacete:2010ad} based on the running coupling BK
(rcBK) equation.

\begin{figure}[ht]
\begin{center}
\includegraphics[width=10cm]{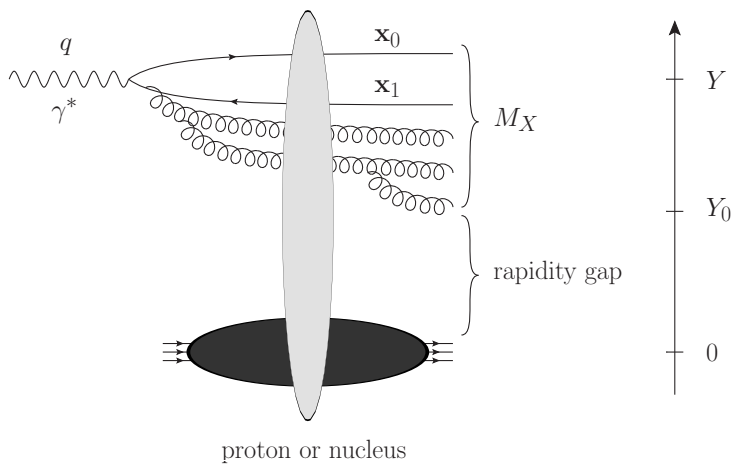}
\end{center}
\caption{Single diffractive dissociation in high energy DIS.}
\label{diff_diss}
\end{figure}

The existing phenomenological applications of the running coupling
evolution to DIS mainly concentrate on the total DIS cross section and
the structure functions \cite{Albacete:2009fh,Albacete:2010sy}. A
related DIS process is the single diffractive dissociation, in which
the proton or nuclear target remains intact in the interaction,
generating an adjacent rapidity gap, with the incoming virtual photon
splitting into a $q\bar q$ pair along with several more gluons and
quarks, which later fragment into hadrons with the net invariant mass
$M_X^2$.  The process is illustrated in \fig{diff_diss}. At very high
energy/small-$x$ the cross section of single diffractive dissociation
is described by the non-linear evolution equation derived in
\cite{Kovchegov:1999ji} in the leading-$\ln 1/x$ and large-$N_c$
approximations with fixed coupling. The equation was generalized
beyond the large-$N_c$ limit in
\cite{Hentschinski:2005er,Kovner:2006ge,Hatta:2006hs}. At the same
time, until now the running coupling corrections to the equation of
\cite{Kovchegov:1999ji} have not been constructed. Similar to the
running coupling corrections for other small-$x$ observables, these
corrections are needed first of all on the purely theoretical grounds,
to determine the scale of the running coupling constant in the kernel
of the equation of \cite{Kovchegov:1999ji}. Second of all, they may be
useful phenomenologically to analyze single diffractive dissociation
in the same framework as the total DIS cross sections. (Note though
that since the invariant mass $M_X^2$ of the hadrons produced in the
single diffractive dissociation at HERA is not very large, in
practical phenomenological applications one often limits the virtual
photon breakup to the $q\bar q$ pair and at most one gluon
\cite{Golec-Biernat:1998js,Buchmuller:1996xw,Kuokkanen:2011je}, thus
alleviating the need for an evolution equation. We will proceed here
expecting that such approximation will require small-$x$ evolution
corrections at the future DIS machines.) The goal of this work is to
find the running coupling corrections to the evolution equation for
diffractive dissociation constructed in \cite{Kovchegov:1999ji}.

This letter is structured as follows. In Sec.~\ref{rederive} we
re-derive the non-linear evolution for the single diffractive
dissociation at fixed-coupling using a method different from the
original one used in \cite{Kovchegov:1999ji}, more similar to the one
used in \cite{Hatta:2006hs}. This new method makes inclusion of the
running coupling corrections straightforward, which is accomplished in
Sec.~\ref{rc}. Our final result is shown in \peq{rcNDevol}, which,
along with the kernels \peq{kbal} and \peq{kkw} demonstrates that the
running coupling corrections to the kernel of the non-linear evolution
equation for diffraction are exactly the same as those for the rcBK
and rcJIMWLK evolution equations.


\section{Re-deriving the fixed-coupling evolution for single 
  diffractive dissociation}
\label{rederive}

We begin by re-deriving the fixed-coupling evolution equation for the
single diffractive dissociation in DIS originally found in
\cite{Kovchegov:1999ji}. Our goal is to find the cross section
$\sigma_{diff}^{\gamma^* A}$ for the single diffractive dissociation
in DIS on a nucleus $A$ resulting in a rapidity gap and production of
hadrons with the invariant mass $M_X^2$, as shown in \fig{diff_diss}.
We denote the net rapidity interval between the projectile $q\bar q$
pair and the target nucleus by $Y = \ln s/Q^2$, where $s$ is the
center-of-mass energy of the $q\bar q$ dipole--nucleus system and $Q^2
= - q^2$ is the photon virtuality. As shown in \fig{diff_diss}, if we
assume that the nucleus has rapidity $0$, then the projectile $q\bar
q$ dipole would have rapidity $Y$. The rapidity gap stretches from
rapidity $0$ to rapidity $Y_0$. Assuming that $s \gg M_X^2 \gg Q^2$
one has $Y_0 \approx \ln s/M_X^2$.

We are working in the kinematics when both $Y_0$ and $Y - Y_0$ are
very large, such that $\as \, Y_0 \sim 1$ and $\as \, (Y-Y_0) \sim 1$
and the leading-logarithmic small-$x$ evolution corrections resumming
powers of $\as \, Y_0$ and $\as \, (Y-Y_0)$ are important. That is, we
need to resum gluon emissions and virtual corrections for the gluons
inside the rapidity gap and in the dissociation region of the $q\bar
q$ pair leading to creation of hadrons with the invariant mass
$M_X^2$.

Following \cite{Kovchegov:1999ji} we denote by $N^D_{{\bf x}_0, {\bf
    x}_1} (Y, Y_0)$ the cross section per unit impact parameter of the
single diffractive dissociation in the dipole--nucleus scattering with
the rapidity gap greater than or equal to $[0, Y_0]$ and the net
rapidity interval of $Y$. As usual ${\bf x}_0$ and ${\bf x}_1$ are the
transverse positions of the quark and the anti-quark in the $q\bar q$
dipole, as shown in \fig{diff_diss}.  (Boldface notation denotes
two-component vectors in the transverse plane.) With the help of $N^D$
the differential cross section for the single diffractive dissociation
in DIS can be written (for a fixed $s$) as
\begin{align}\label{xsec}
  M_X^2 \, \frac{d \sigma_{diff}^{\gamma^* A}}{d M_X^2} = - \int d^2
  x_0 \, d^2 x_1 \int\limits_0^1 \, dz \ |\Psi^{\gamma^* \rightarrow q
    {\bar q}} ({x}_{01}, z)|^2 \, \frac{\pd N^D_{{\bf x}_0, {\bf x}_1}
    (Y, Y_0)}{\pd Y_0},
\end{align}
with $|\Psi^{\gamma^* \rightarrow q {\bar q}} ({x}_{01}, z)|^2$ the
standard order-$\alpha_{EM}$ light-cone wave function squared for a
virtual photon fluctuating into a $q {\bar q}$ pair with $x_{01} =
|{\bf x}_0 - {\bf x}_1|$ (see e.g.  \cite{Kovchegov:1999yj}).

We want to write down an evolution equation for $N^D_{{\bf x}_0, {\bf
    x}_1} (Y, Y_0)$. First we note that when $Y_0 = Y$ the rapidity
gap spans the whole rapidity interval and the interaction is elastic.
In such case \cite{Kovchegov:1999ji}
\begin{align}\label{initN}
  N^D_{{\bf x}_0, {\bf x}_1} (Y = Y_0, Y_0) = [ N_{{\bf x}_0, {\bf
      x}_1} (Y_0)]^2
\end{align}
where $N_{{\bf x}_0, {\bf x}_1} (Y)$ is the dipole--nucleus forward
scattering amplitude defined by
\begin{align}\label{Ndef}
  N_{{\bf x}_0, {\bf x}_1} (Y) = \frac{1}{N_c} \, \left\langle
    \mbox{tr} \left[ 1 - V_{{\bf x}_0} \, V^\dagger_{{\bf x}_1}
    \right] \right\rangle_Y
\end{align}
with $V_{\bf x}$ the Wilson line describing the interaction of a quark
at $\bf x$ moving along the $x^+$ axis with the target
\begin{align}\label{Vlimits_def}
  V_{\bf x} = \mbox{P} \exp \left\{ i \, g \,
    \int\limits_{-\infty}^{\infty} d x^+ \, t^a \, A^{a \, -} (x^+,
    x^- =0, {\bf x}) \right\}.
\end{align}
Here $A^\mu$ is the operator of the gluon field of target, $t^a$ are
fundamental generators of SU($N_c$), and $\langle \ldots \rangle_Y$ in
\eq{Ndef} denotes averaging in the target wave function evolved up to
rapidity $Y$ (see \cite{Iancu:2003xm,Weigert:2005us} for reviews of
this notation). The light cone coordinates are defined as $x^\pm = (t
\pm z)/\sqrt{2}$.

In the large-$N_c$ limit and in the leading-$\ln s$ approximation
$N_{{\bf x}_0, {\bf x}_1} (Y)$ obeys the BK evolution equation
\cite{Balitsky:1996ub,Balitsky:1997mk,Balitsky:1998ya,Kovchegov:1999yj,
  Kovchegov:1999ua}
\begin{align}\label{BKeqn}
  \pd_Y N_{{\bf x}_0, {\bf x}_1} (Y) = \frac{\as \, N_c}{2 \, \pi^2}
  \, \int d^2 x_2 \, \frac{x_{01}^2}{x_{02}^2 \, x_{21}^2} \, \left[
    N_{{\bf x}_0, {\bf x}_2} (Y) + N_{{\bf x}_2, {\bf x}_1} (Y) -
    N_{{\bf x}_0, {\bf x}_1} (Y) \right. \notag \\ \left. - N_{{\bf
        x}_0, {\bf x}_2} (Y) \, N_{{\bf x}_2, {\bf x}_1} (Y) \right]
\end{align}
where $x_{ij} = |{\bf x}_{ij}|$ with ${\bf x}_{ij} = {\bf x}_i - {\bf
  x}_j$ and $\pd_Y = \partial/\partial Y$. The solution of \eq{BKeqn}
would allow us to find the elastic dipole--nucleus scattering cross
section (per unit impact parameter) via \eq{initN}. The resulting
$N^D_{{\bf x}_0, {\bf x}_1} (Y = Y_0, Y_0)$ would serve as the initial
condition for the $Y$-evolution of $N^D_{{\bf x}_0, {\bf x}_1} (Y,
Y_0)$ we are about to construct.

\begin{figure}[ht]
\begin{center}
\includegraphics[width=\textwidth]{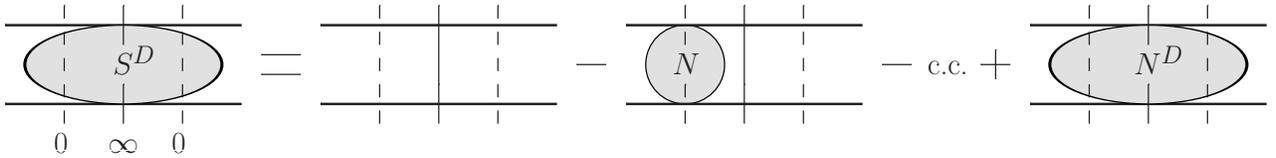}
\end{center}
\caption{Diagrammatic representation of $S^D$ defined in \eq{SD_def}.}
\label{Sdef}
\end{figure}

Note that $N^D$ is a scattering cross section (per unit impact
parameter), and, therefore, when written as amplitude squared,
contains interactions to the left and to the right of the final state
cut. (For example, in \eq{initN} the interactions are contained in the
two factors of $N$.)  It is easier to derive its evolution by
completing it to the $S$-matrix like object by defining
\begin{align}\label{SD_def}
  S^D_{{\bf x}_0, {\bf x}_1} (Y, Y_0) = 1 - 2 \, N_{{\bf x}_{0}, {\bf
      x}_{1}} (Y) + N^D_{{\bf x}_{0}, {\bf x}_{1}} (Y, Y_0).
\end{align}
The new quantity $S^D$ contains the no-interaction term ($=1$), and
two terms with the interactions only to the left or to the right of
the cut (giving $-N_{{\bf x}_{0}, {\bf x}_{1}} (Y)$ each), in addition
to $N^D$. The addition of these extra terms which have no new
particles produced in the final state to $N^D$ means that all terms in
$S^D$ describe the final state with the rapidity gap greater than or
equal to $Y_0$, allowing for no-interaction contributions to the right
and to the left of the cut. The quantity $S^D$ is illustrated in
\fig{Sdef}, where
the solid vertical straight line denotes the final state cut at the
light-cone time $x^+ = \infty$, while the dashed vertical straight
lines denote the Glauber-Mueller interaction with the target
\cite{Mueller:1989st} which happens at the time $x^+ =0$ both in the
amplitude and in the complex conjugate amplitude. The times are
labeled in the leftmost diagram in \fig{Sdef}. (The time scale of the
interaction with the target is much shorter than the time scales of
the small-$x$ evolution: see
\cite{Jalilian-Marian:2005jf,Weigert:2005us,Iancu:2003xm} for reviews
of similar calculations.) Shaded ovals and the circle in \fig{Sdef}
represent all gluon emissions and interactions with the target nucleus
contained in $S^D$, $N$, and $N^D$ as labeled.

\begin{figure}[ht]
\begin{center}
\includegraphics[width=0.7\textwidth]{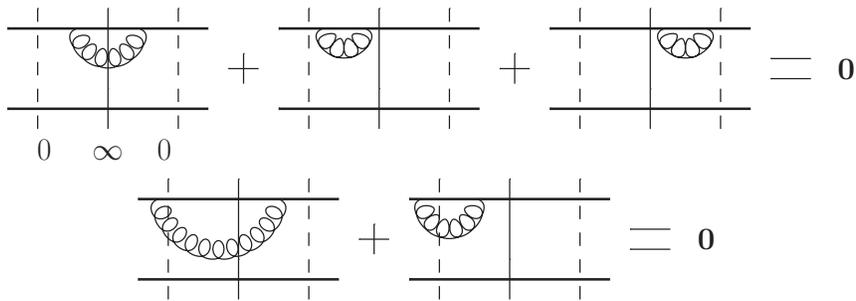}
\end{center}
\caption{Diagrammatic examples of the final state cancellations derived 
  in \cite{Chen:1995pa}. Cancellation for the sum of the diagrams
  which are complex conjugates of those shown in the bottom panel also
  takes place, but is not shown explicitly.}
\label{cancel}
\end{figure}

To construct one step of small-$x$ evolution for $S^D$ we have to see
how this quantity changes under an emission of a gluon with rapidity
$Y > Y_0$. Note that for $Y=Y_0$ \eq{initN} leads to 
\begin{align}\label{initS}
  S^D_{{\bf x}_0, {\bf x}_1} (Y=Y_0, Y_0) = \left[ 1 - N_{{\bf x}_{0},
      {\bf x}_{1}} (Y_0) \right]^2,
\end{align}
which is the initial condition for the evolution of $S^D$. In one step
of small-$x$ evolution a soft gluon can be emitted and absorbed by the
quark and the anti-quark in the dipole $01$ at any time $x^+ \in
(-\infty, +\infty)$ both in the amplitude and in the complex conjugate
amplitude. The rapidity gap constraint on the final state of $S^D$
implies that no gluon should be present at $x^+ = \infty$ for gluon
rapidities less than $Y_0$. However, this condition is already
satisfied by our initial condition \peq{initS}. There are no
final-state restrictions on the gluons with $Y > Y_0$: they may or may
not go through the cut. This makes the gluon emissions and absorptions
at times $x^+ \in (0, +\infty)$ in the amplitude and in the complex
conjugate amplitude cancel, due to the final-state cancellations
originally derived in \cite{Chen:1995pa} (see e.g.
\cite{Kovchegov:1999ji,Kovchegov:2001sc} for applications of such
cancellations). The final state cancellations of \cite{Chen:1995pa}
are illustrated in \fig{cancel} which shows how the diagrams with the
gluon either emitted or absorbed (or both) at $x^+ \in (0, +\infty)$
cancel. (Cancellations also take place for all other couplings of the
gluon to the dipole not shown explicitly in \fig{cancel}.)

\begin{figure}[ht]
\begin{center}
\includegraphics[width=0.8\textwidth]{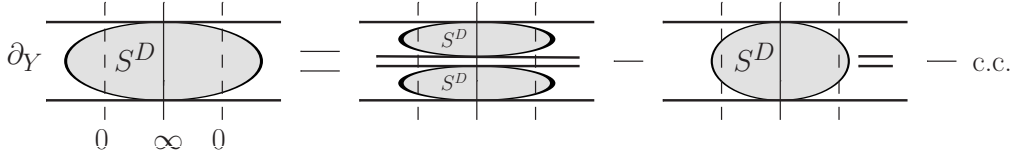}
\end{center}
\caption{The evolution equation for $S^D$.}
\label{SDevol_fig}
\end{figure}

This leaves us with gluon emissions/absorptions at $x^+ \in (-\infty,
0)$ both in the amplitude and in the complex conjugate amplitude.
Identifying the time interval $x^+ \in (-\infty, 0)$ in the complex
conjugate amplitude with the time interval $x^+ \in (0, +\infty)$ in
the forward amplitude we conclude that the evolution for $S^D$ with
$Y>Y_0$ is the same as for the $S$-matrix of the dipole--nucleus
scattering, which is related to the forward amplitude $N$ by 
\begin{align}
  \label{S-matrix}
  S_{{\bf x}_{0}, {\bf x}_{1}} (Y) = 1 - N_{{\bf x}_{0}, {\bf x}_{1}}
  (Y).
\end{align}
In the large-$N_c$ limit it is described by the BK equation
\peq{BKeqn} re-written for the $S$-matrix:
\begin{align}\label{SDevol}
  \pd_Y S^D_{{\bf x}_{0}, {\bf x}_{1}} (Y, Y_0) \, = \, \frac{\as \,
    N_c}{2 \, \pi^2} \, \int \, d^2 x_2 \,
  \frac{x^2_{10}}{x^2_{20}\,x^2_{21}} \, \left[ S^D_{{\bf x}_{0}, {\bf
        x}_{2}} (Y, Y_0) \, S^D_{{\bf x}_{2}, {\bf x}_{1}} (Y, Y_0) -
    S^D_{{\bf x}_{0}, {\bf x}_{1}} (Y, Y_0) \right].
\end{align}
The evolution equation \peq{SDevol} is illustrated in
\fig{SDevol_fig}, where, in the large-$N_c$ approximation notation
gluons are denoted by the double lines and disconnected gluon (double)
line implies all possible connections to the quark and the anti-quark
in the dipole (see
\cite{Mueller:1994rr,Mueller:1994jq,Mueller:1995gb}).

One could stop at \eq{SDevol}, since its solution with the initial
condition \peq{initS} used in \eq{SD_def} would give one $N_D$. (Even
more, \eq{SD_def} implies that $\pd_{Y_0} S^D = \pd_{Y_0} N^D$, which
is all that is needed to find the differential diffractive cross
section in \eq{xsec}.) Alternatively one can use \eq{SD_def} along
with \eq{BKeqn} in \eq{SDevol} to obtain the evolution equation for
$N^D$ \cite{Kovchegov:1999ji}
\begin{align}\label{NDevol}
  \pd_Y N^D_{{\bf x}_{0}, {\bf x}_{1}} (Y, Y_0) \, = \, \frac{\as \,
    N_c}{2 \, \pi^2} \, \int \, d^2 x_2 \,
  \frac{x^2_{10}}{x^2_{20}\,x^2_{21}} \bigg[ N^D_{{\bf x}_{0}, {\bf
      x}_{2}} (Y, Y_0) + N^D_{{\bf x}_{2}, {\bf x}_{1}} (Y, Y_0) -
  N^D_{{\bf x}_{0}, {\bf x}_{1}} (Y, Y_0) \notag \\ + N^D_{{\bf
      x}_{0}, {\bf x}_{2}} (Y, Y_0) \, N^D_{{\bf x}_{2}, {\bf x}_{1}}
  (Y, Y_0) - 2 \, N_{{\bf x}_{0}, {\bf x}_{2}} (Y) \, N^D_{{\bf
      x}_{2}, {\bf x}_{1}} (Y, Y_0) - 2 \, N^D_{{\bf x}_{0}, {\bf
      x}_{2}} (Y, Y_0) \, N_{{\bf x}_{2}, {\bf x}_{1}} (Y) \notag \\ +
  \, 2 \, N_{{\bf x}_{0}, {\bf x}_{2}} (Y) \, N_{{\bf x}_{2}, {\bf
      x}_{1}} (Y) \bigg].
\end{align}


\section{Running coupling corrections}
\label{rc}

Similar to the previous works on the running coupling corrections to
small-$x$ observables
\cite{Kovchegov:2006vj,Balitsky:2006wa,Gardi:2006rp,Albacete:2007yr,Kovchegov:2007vf,Horowitz:2010yg}
we use the Brodsky--Lepage--Mackenzie (BLM) prescription
\cite{Brodsky:1983gc}: one has to dress all the gluon propagators and
multi-gluon vertices with quark bubble corrections, and then replace
the number of quark flavors $N_f$ by
\begin{align}\label{repl}
N_f \rightarrow - 6 \, \pi \, \beta_2
\end{align}
where 
\begin{align}\label{beta}
\beta_2 = \frac{11 N_c - 2 N_f}{12 \, \pi}  
\end{align} 
is the one-loop QCD beta-function. In the end one is able to absorb
all such corrections into the powers of the physical running couplings
\begin{align}\label{as_geom}
  \as (Q^2) \, = \, \frac{\amu}{1 + \amu \, \beta_2 \, \ln
    \frac{Q^2}{\mu^2}}.
\end{align}
This procedure fixes the scales $Q$ in the running couplings.

When the gluon lines in our above derivation of \eq{NDevol} are
dressed by the quark loop corrections, the final state cancellations
of \cite{Chen:1995pa} shown in \fig{cancel} still take place. All the
$g \to q \bar q$ splitting and the $q {\bar q} \to g$ mergers
happening at the times $x^+ \in (0, +\infty)$ in the amplitude and in
the complex conjugate amplitude cancel (simultaneously with the
cancellation of gluon emissions and absorptions by the original dipole
at $x^+ \in (0, +\infty)$ shown in \fig{cancel}). These cancellations
have been discussed in detail in \cite{Kovchegov:2007vf}. An example
of a cancellation of the final state interactions with a quark loop
correction added to the top line of \fig{cancel} is shown in
\fig{rc_fs}: as we demonstrate in Appendix A, the sum of all the
diagrams shown in \fig{rc_fs} is zero,
\begin{align}
  A + B + C + D + E =0. 
\end{align}
The argument can be repeated at higher orders, showing that higher
numbers of quark loops would also cancel in the final state, and for
other diagrams with the final-state interactions.

\begin{figure}[ht]
\begin{center}
\includegraphics[width=\textwidth]{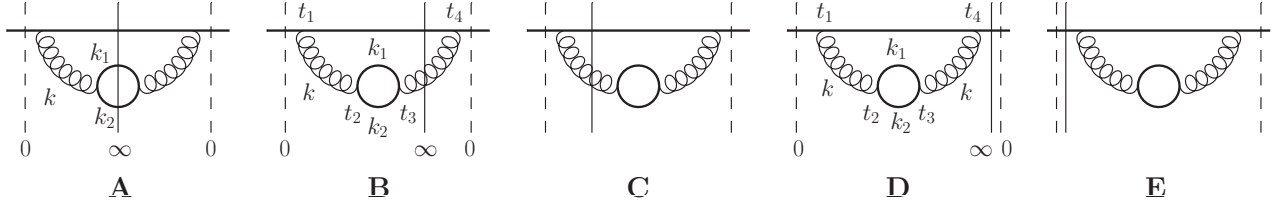}
\end{center}
\caption{An example of the final-state interactions with a quark loop 
  correction: the sum of these diagrams is zero, as shown in Appendix
  A.}
\label{rc_fs}
\end{figure}

The remaining $g \to q \bar q$ splitting and $q {\bar q} \to g$
mergers are limited to the $x^+ \in (-\infty, 0)$ times both in the
amplitude and in the complex conjugate amplitude, thus dressing the
propagators of the early-time gluons only. Therefore, the running
coupling evolution for $S^D$ is the same as the running coupling
evolution for the BK equation \peq{BKeqn} written for the
dipole--nucleus $S$-matrix of \eq{S-matrix}. Using the results of the
derivation of the rcBK evolution in
\cite{Kovchegov:2006vj,Balitsky:2006wa} we write the running coupling
evolution equation for $S^D$ as
\begin{align}\label{SDevol_rc}
  \pd_Y S^D_{{\bf x}_{0}, {\bf x}_{1}} (Y, Y_0) \, = \, \int \, d^2
  x_2 \, K ({\bf x}_0, {\bf x}_1, {\bf x}_2) \, \left[ S^D_{{\bf
        x}_{0}, {\bf x}_{2}} (Y, Y_0) \, S^D_{{\bf x}_{2}, {\bf
        x}_{1}} (Y, Y_0) - S^D_{{\bf x}_{0}, {\bf x}_{1}} (Y, Y_0)
  \right].
\end{align}
The kernel $K ({\bf x}_0, {\bf x}_1, {\bf x}_2)$ of the evolution
equation \peq{SDevol_rc} depends on the prescription used to fix the
scales of the running couplings (see \cite{Albacete:2007yr} for
details). In the Balitsky prescription \cite{Balitsky:2006wa} the
kernel is
\begin{align}\label{kbal}
  K_{rc}^{Bal} ({\bf x}_{0}, {\bf x}_{1}, {\bf x}_{2}) = \frac{N_c \,
    \alpha_s (x_{10}^2)}{2\pi^2} \Bigg[
  \frac{x^2_{10}}{x^2_{20}\,x^2_{21}} +
  \frac{1}{x_{20}^2}\left(\frac{\alpha_s(x_{20}^2)}{\alpha_s(x_{21}^2)}-1\right)+
  \frac{1}{x_{21}^2}\left(\frac{\alpha_s(x_{21}^2)}{\alpha_s(x_{20}^2)}-1\right)
  \Bigg],
\end{align}
with the following shorthand notation
\begin{align}
  \alpha_s (x_\perp^2) = \as \left( \frac{4 \, e^{-\frac{5}{3} - 2 \,
        \gamma_E}}{x_\perp^2} \right)
\end{align}
in the $\overline{\text{MS}}$ renormalization scheme. In the
Kovchegov--Weigert prescription \cite{Kovchegov:2006vj} the kernel is
\begin{align}\label{kkw}
  K_{rc}^{KW} ({\bf x}_{0}, {\bf x}_{1}, {\bf x}_{2}) =
  \frac{N_c}{2\pi^2} \, \Bigg[ \alpha_s(x_{20}^2)\frac{1}{x_{20}^2}-
  2\,\frac{\alpha_s(x_{20}^2)\,\alpha_s(x_{21}^2)}{\alpha_s
    (R^2)}\,\frac{ {\bf x}_{20} \cdot {\bf
      x}_{21}}{x_{20}^2\,x_{21}^2} +
  \alpha_s(x_{21}^2)\frac{1}{x_{21}^2} \Bigg]
\end{align}
with 
\begin{align}
  R^2 = x_{20} \, x_{21} \left(\frac{x_{21}}{x_{20}}\right)^
  {\frac{x_{20}^2+x_{21}^2}{x_{20}^2-x_{21}^2}-2\,\frac{x_{20}^2\,x_{21}^2}{
      {\bf x}_{20} \cdot {\bf x}_{21}}\frac{1}{x_{20}^2-x_{21}^2}}.
\end{align}
The terms in the kernel neglected by each prescription can be found in
\cite{Albacete:2007yr,Balitsky:2006wa}.

The initial condition for the running coupling evolution for $S^D$ is
given by \eq{initS}, where $N$ is now found from the rcBK equation
\cite{Kovchegov:2006vj,Balitsky:2006wa}
\begin{align}\label{rcBKeqn}
  \pd_Y N_{{\bf x}_0, {\bf x}_1} (Y) = \int d^2 x_2 \, K ({\bf x}_0,
  {\bf x}_1, {\bf x}_2) \, \left[ N_{{\bf x}_0, {\bf x}_2} (Y) +
    N_{{\bf x}_2, {\bf x}_1} (Y) - N_{{\bf x}_0, {\bf x}_1} (Y)
  \right. \notag \\ \left. - N_{{\bf x}_0, {\bf x}_2} (Y) \, N_{{\bf
        x}_2, {\bf x}_1} (Y) \right].
\end{align}

Finally, using \eq{SD_def} in \eq{SDevol_rc} one obtains the
running coupling evolution for $N^D$
\begin{align}\label{rcNDevol}
  \pd_Y N^D_{{\bf x}_{0}, {\bf x}_{1}} (Y, Y_0) \, = \, \int \, d^2
  x_2 \, K ({\bf x}_0, {\bf x}_1, {\bf x}_2) \, \bigg[ N^D_{{\bf
      x}_{0}, {\bf x}_{2}} (Y, Y_0) + N^D_{{\bf x}_{2}, {\bf x}_{1}}
  (Y, Y_0) - N^D_{{\bf x}_{0}, {\bf x}_{1}} (Y, Y_0) \notag \\ +
  N^D_{{\bf x}_{0}, {\bf x}_{2}} (Y, Y_0) \, N^D_{{\bf x}_{2}, {\bf
      x}_{1}} (Y, Y_0) - 2 \, N_{{\bf x}_{0}, {\bf x}_{2}} (Y) \,
  N^D_{{\bf x}_{2}, {\bf x}_{1}} (Y, Y_0) - 2 \, N^D_{{\bf x}_{0},
    {\bf x}_{2}} (Y, Y_0) \, N_{{\bf x}_{2}, {\bf x}_{1}} (Y) \notag
  \\ + \, 2 \, N_{{\bf x}_{0}, {\bf x}_{2}} (Y) \, N_{{\bf x}_{2},
    {\bf x}_{1}} (Y) \bigg]
\end{align}
with the kernel given by \eq{kbal} or by \eq{kkw} depending on the
prescription choice and with the initial condition given by \eq{initN}
with $N$ provided by \eq{rcBKeqn}.

Eqs.~\peq{SDevol_rc} and \peq{rcNDevol} are the main results of this
letter, providing running coupling corrections to the non-linear
small-$x$ evolution for the cross section of diffractive dissociation
in DIS. The kernels in Eqs.~\peq{kbal} and \peq{kkw} have of course
been found before in the calculation of the rcBK and rcJIMWLK
evolutions \cite{Kovchegov:2006vj,Balitsky:2006wa}: we showed that
these same kernels govern the evolution for diffractive dissociation.
Our derivation was done in the large-$N_c$ limit: since the evolution
equation \peq{SDevol_rc} is simply the rcBK equation, its
generalization to all-$N_c$ is the same as for the rcBK equation, and
is accomplished by performing the $\langle \ldots \rangle_{Y, Y_0}$
averaging to the whole right-hand side of \eq{SDevol_rc} instead of
applying it to each individual $S^D$ separately (see
\cite{Weigert:2000gi,Kovner:2000pt}). Note that the averaging should
now include the final-state rapidity gap greater than or equal to $[0,
Y_0]$.


\section{Acknowledgments}

The author would like to thank Genya Levin and Heribert Weigert for
discussions on the subject.

This research is sponsored in part by the U.S. Department of Energy
under Grant No. DE-SC0004286.


\appendix

\renewcommand{\theequation}{A\arabic{equation}}
  \setcounter{equation}{0}
\section{Final state cancellations}
\label{A}

We want to demonstrate the cancellation of diagrams in \fig{rc_fs}. We
will be working in the light-cone perturbation theory (LCPT)
\cite{Lepage:1980fj,Brodsky:1997de}. In LCPT the diagrams in
\fig{rc_fs} differ only by their energy denominators and overall
signs. 
Suppressing the rest of the diagrams we write for the contribution of
the diagram $A$
\begin{align}\label{Acontr}
  A = \frac{1}{(E_1 + E_2)^2 \, (E_1 + E_2 - E_k)^2}
\end{align}
where $E_k = {\bf k}^2/ (2 \, k^+)$ is the light cone energy of a line
with momentum $k$, and $E_1 = {\bf k}_1^2/ (2 \, k_1^+)$, $E_2 = {\bf
  k}_2^2/ (2 \, k_2^+)$ (momentum labels are explained in
\fig{rc_fs}).

The remaining diagrams have to be treated more carefully, due to the
presence of intermediate states with zero light-cone energy
denominators \cite{Chen:1995pa,Kovchegov:2007vf}. Noting that the
diagram $B$ has an overall minus sign compared to $A$ due to moving a
quark-gluon vertex across the cut we write
\begin{align}
  \label{Bcontr}
  B & = - \int\limits_0^\infty d t_3 \, e^{i \, (E_k - E_1 - E_2 + i
    \, \epsilon) \, t_3} \, \int\limits_0^{t_3} d t_2 \, e^{i \, (E_1
    + E_2 - E_k + i \, \epsilon) \, t_2} \, \int\limits_0^{t_2} d t_1
  \, e^{i \, (E_k + i \, \epsilon) \, t_1} \, \int\limits_0^\infty d
  t_4 \, e^{-i \, (E_k - i \, \epsilon) \, t_4} \notag \\ & =
  \frac{i}{2 \, \epsilon \, E_k^2 \, (E_1 + E_2 - E_k)} + \frac{2 \,
    E_1 + 2 \, E_2 - 3 \, E_k}{2 \, E_k^3 \, (E_1 + E_2 - E_k)^2} + O
  (\epsilon),
\end{align}
where all the light cone times associated with the vertices, $t_i =
x^+_i$, are labeled in \fig{rc_fs}B and $\epsilon$ is an infinitesimal
regulator. (See \cite{Chen:1995pa,Kovchegov:2007vf} for similar
calculations.) One can show that the singular $O(1/\epsilon)$ term in
\eq{Bcontr} is canceled by the diagrams with the gluon lines replaced
by the instantaneous interactions of LCPT: we thus concentrate on the
$O(\epsilon^0)$ term. Also, due to mirror symmetry, $C = B^*$, such
that $O (1/\epsilon)$ terms would cancel in the $B+C$ sum
\cite{Chen:1995pa}.

Similarly for the diagrams $D$ and $E$ we get (see \fig{rc_fs}D for
notations)
\begin{align}
  \label{Dcontr}
  D & = \int\limits_0^\infty d t_4 \, e^{i \, (- E_k + i \, \epsilon)
    \, t_4} \, \int\limits_0^{t_4} d t_3 \, e^{i \, (E_k - E_1 - E_2 +
    i \, \epsilon) \, t_3} \, \int\limits_0^{t_3} d t_2 \, e^{i \,
    (E_1 + E_2 - E_k + i \, \epsilon) \, t_2} \, \int\limits_0^{t_2} d
  t_1 \, e^{i \, (E_k + i \, \epsilon) \, t_1} \notag \\ & =
  \frac{i}{4 \, \epsilon \, E_k^2 \, (E_1 + E_2)} - \frac{2 \, E_1 + 2
    \, E_2 + E_k}{2 \, E_k^3 \, (E_1 + E_2)^2} + O (\epsilon).
\end{align}
Again, due to mirror symmetry, $E = D^*$. With the help of
Eqs.~\peq{Acontr}, \peq{Bcontr}, and \peq{Dcontr} and dropping the $O
(\epsilon)$ terms we see that
\begin{align}
  A + B + C + D + E = 0,
\end{align}
as desired.



\providecommand{\href}[2]{#2}\begingroup\raggedright\endgroup


\end{document}